# Device-Life Boundary Indicators and the Classification of Restored Versus Native Records on Android

Daniel Kwaku Ntiamoah Addai
SectraTEK Lab, Prudential Associates
daniel@prudentialassociates.com
*Corresponding author*

Selorm Adjabeng
Cyberteq Falcon Ltd
sirlormselorm@gmail.com

Kingsley Albert Egyir
Northeastern University
egyir.k@northeastern.edu

Sanders Larweh Tetteh
Booking.com
sanderstetteh@gmail.com

**ABSTRACT**

A phone that shows calls, messages and photographs from years ago may have been set up recently and carried them over from an earlier handset, so its screen dates do not date the device's own life. Which traces fix where that life begins, how well they agree and how much of what is visible predates it is unmeasured. We read 14 Android full file system extractions (12 Samsung; Android 10 to 16), seven bug reports and 33 iOS extractions (a probe). Every Android device recorded a first boot, but the first evidence of user setup came a median of 98 days later (0 to 563). Of the 13 devices with messages, 7 held some dated before the first boot, a majority of them on 5; 24% of the apps outside the system image were installed as restored. Against the creator field that a migration tool leaves on messages, a boundary at user setup recalled all but 1 of 46,968 such messages, the first-boot boundary 94 to 100%. On synthetic fixtures, the labelling error followed the boundary error. A date on a record dates the device only once the record is shown to have been created on it.

## 1 INTRODUCTION

A date on a phone is evidence about a day only if the phone held that record on that day. Users change handsets and bring their data with them: a cloud restore or a device-to-device transfer writes years of calls, messages and photographs into a handset that may be weeks old. The new device then shows records that predate it. Some stores keep the original dates of what they receive, others stamp what arrives with the time of arrival, and the examiner sees both side by side. A report that reads the oldest message on a phone as the start of that phone's use, or an app's install time as the day its user first used the app, can be wrong by years.

**A motivating example.** On the 13 devices in our sample that held a message store, 161,521 of 226,082 messages (71%) are dated before the first boot recorded on the same device, and on 5 of those devices they are the majority. Of 17,971 calls on 13 devices, 24% are dated before it. Of the apps outside the system image on the 14 devices, 24% carry the install reason "restored from another device" (Section 5.3). Records dated before the first boot were not created on the device holding them, unless its clock was wrong or the writing app set the date.

**The problem.** Work on reset artifacts dates the last wipe of a device [1], [14], [16], and studies of restores show that restored data and timestamps differ from the original [11], [12]. What is missing is the step between them: which traces on an Android phone fix the start of its own life, how closely they agree, what share of the records an examiner sees lies before that start, and whether a rule can label a record as created on the device or carried to it, with a stated error. We ask five research questions.

- **RQ1 (Indicators).** Which independent traces fix the start of a device's own life, and how often do they exist?
- **RQ2 (Agreement).** How closely do they agree?
- **RQ3 (Predating records).** What share of the visible calls, messages, media and app installs predates the boundary?
- **RQ4 (Labelling).** Can a rule label a record native or restored, and with what error?
- **RQ5 (Consequences).** What follows for a claim about a date?

**Contributions.** This report makes four contributions.

- Two defined boundaries, the first-boot record and the first evidence of user setup, with a list of the traces that mark them, the direction in which each can be wrong, and the platform source that documents each (Sections 2 and 3).
- A measurement of how often those traces exist and how far apart they lie on 14 Android extractions, with 7 bug reports and a probe of 33 iOS extractions (Sections 5.1, 5.2 and 5.8).
- The share of calls, messages, media and installs that predates the boundaries, and a check of the boundary rules against the creator field that a migration tool writes (Sections 5.3 to 5.5).
- Four labelling rules with their error on 200 synthetic fixture runs per case built to the documented write behaviour, and a ladder for wording a date claim (Sections 5.7 and 6).

**Table 1:** Research questions, answers and evidence. Counts are devices of the 14 Android extractions unless stated.

| RQ | Answer | Evidence |
|---|---|---|
| RQ1 | Every device recorded a first boot (14 of 14); user setup is evidenced by four upper-bound traces, on 10 to 14 of 14 devices; a backup record, a setup timestamp file and a message creator field exist on a minority (3, 3, 3 devices). | Table 4 |
| RQ2 | First boot and first user setup are two events: setup came a median of 98 days later (0 to 563); 7 of 14 devices at least 100 days later, 2 within a day. The first media index agrees with the first boot on 12 of 14 devices. | Table 5, Fig. 2 |
| RQ3 | Messages: 7 of 13 devices hold some before the first boot, 5 a majority. Calls: 5 and 3. Photograph and video capture dates: 10 and 3 of 14. 24% of installs outside the system image are marked restored. | Table 6, Fig. 3 |
| RQ4 | Against the creator field, a user-setup boundary recalls 46,967 of 46,968 migration-tool messages, the first-boot boundary 44,650; the order rule, which needs no boundary, 46,741 with 0 other messages flagged. On fixtures with an exact boundary the combined rule recalls at least 99.98% of restored records in every scenario. | Tables 8 and 9, Fig. 4 |
| RQ5 | A record dated before the first boot cannot be a record of this device, if the clock was right at that boot; between first boot and user setup it is doubtful; only after user setup does its date date the device. | Section 6 |

**Scope.** The Android results describe the Lab's research corpus of Android phones: 14 full file system extractions readable when the study ran, 12 of them Samsung, Android 10 to 16, most Android 16. The analysis read stored extractions and bug reports only and never connected to a device; every finding is read from stored data, and no record of how or when a device was restored was used. The creator-field check covers three Samsung devices that used one migration tool. The iOS part is a probe on undocumented marker files. Section 7 states the limits.

## 2 BACKGROUND AND RELATED WORK

This section describes what Android records at the start of a device's life and what a restore carries over, then places the work among studies of reset dating and of restored data.

### 2.1 What Android records, and what a restore carries

When the data partition is empty, the platform's boot-statistics service creates a `factory_reset` event at the first boot-time record and stores its time in the file's modification time; a comment in the code says that the event does not exist after a reset and therefore marks the time of the reset [4]. We call this time the *first boot record*. User setup is recorded differently: a flag says that the setup wizard finished [7], without a time. Times appear in the traces that setup and the first minutes of use create. The package manager stores, per user and package, a first-install time and an install reason [6]; the reason distinguishes a user-initiated install, an install during device setup and an install "as part of restoring from another device" [5]. The backup manager persists a record of the restore set it came from and the set it currently uses [9]. The message provider fills a read-only *creator* column, documented as usually the package name of the app that sends a message [8]; we read the stored value as the package that wrote the row. The message and call tables in the Android providers use `AUTOINCREMENT` row identifiers, which are never reused and only increase [17], so identifier order is write order. The media index keeps the time a file was added and the time it was taken [13]. The user record has a creation time [10].

A restore writes records that were created elsewhere into these stores. What it keeps and what it replaces depends on the store and the tool: the message provider keeps the date it is given, the package manager stamps the install with the time of installation and the reason "restore" [5]. Which of these hold for a given device is the question of Section 5.

### 2.2 Dating the last reset

Schwamm and Rowe studied what remains on mobile devices after a factory reset [16]. Keçe and colleagues reset Android and iOS devices and report that system files such as the setup-wizard preferences and Google service files are recreated, so that their file times agree to the minute with the reset [14]. Our earlier report catalogued the reset-related sources of Android phones and the clocks behind their dates, and found each informative but individually unreliable [1]; the retention of the stores that could reach back to a boot is bounded [2], and clocks that are wrong at the first boot are not rare [3]. These studies date the reset. None asks which of the records on the device predate it.

### 2.3 Restores and timestamps

Geus and colleagues compared a full file system extraction and a local backup of an iOS device with an iCloud backup restored to a test device and report significant discrepancies in timestamp information and minor differences in user data [12]; an earlier study found local backups to be correct copies of the original in most but not all cases [11]. A vendor webinar is described as showing how to find out which backup or restore method moved data between Android devices, on stock Android and on Samsung; its public description lists no artifacts [15]. We found no public study that lists the traces which mark where an Android device's own life begins, measures their agreement across a set of devices, or measures the share of visible records that predates them (search of the ACM Digital Library, DFRWS, *Forensic Science International: Digital Investigation*, arXiv and the web, 28 October 2026).

### 2.4 Position of this work

We do not compare tools and we did not run controlled restores. We measure, on devices as they were acquired, which stored traces bound the start of a device's life, and we test labelling rules against the one stored label that exists, the creator of a message, and against synthetic fixtures. The first-boot record, the install reason, the backup record and the creator column are documented in the platform source; their use as boundary and restore indicators, and the numbers, are new. Comparing rule outputs with the creator field is a check on stored data, the rules and fixtures are engineering, and no part of the work replicates an earlier study.

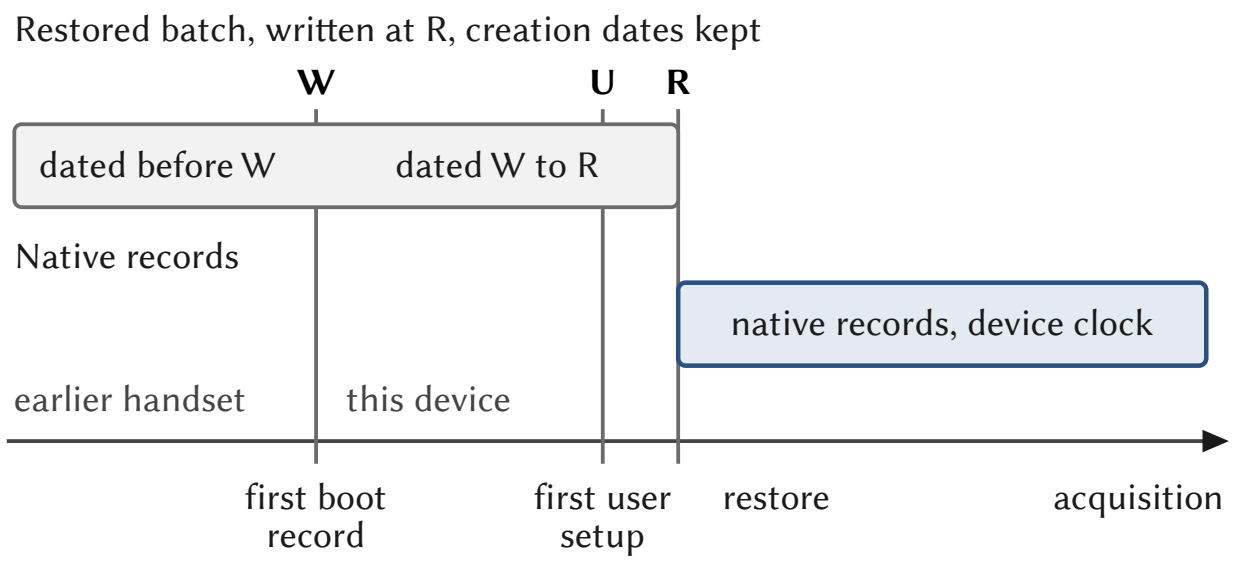

**Figure 1:** The boundaries of Section 3.1 on a device timeline. A restored batch is written in one burst at R and keeps the creation dates it was given; the part dated before W cannot have been created on this device.

## 3 METHOD

We define two boundaries, list the traces that mark them, define record labels and rules, and state how the rules are tested.

### 3.1 Boundaries and labels

The *first boot boundary W* is the first boot record of the data partition (Section 2.1). A record created on the device cannot be dated before W if the device clock was right at that boot; we test that in Section 5.2. The *user setup boundary U* is the earliest evidence that a user set the device up. A device may sit in a store, on a shelf or at a carrier between W and U, so U is at or after W. A record is *native* if the device created it on its own clock, *restored* if a backup, transfer or sync wrote it from elsewhere, and *carried* if either is possible, for example a file received from another phone. We use *predating* for a record whose date is more than one hour before a boundary.

**Table 2:** Candidate traces for the start of a device's life: what each marks, how it can err, and where the platform documents it. "Observed" means we found the trace in extractions and have no platform documentation for it.

| Trace | What it is, what it marks, how it can err | Ref. |
|---|---|---|
| First boot record | File time of the `factory_reset` event in the boot-statistics directory. Marks: W. Can err: wrong if the clock was wrong at that boot; reset by any later wipe. | [4] |
| First media index | Earliest `date_added` in the media store. Marks: about W. Can err: the media provider scans at first boot; later if its database was rebuilt. | [13] |
| Account add log | Earliest `action_account_add` in the account database log. Marks: U upper bound. Can err: retains the last 64 actions only. | observed |
| Account password time | Oldest last-password-entry time over accounts. Marks: U upper bound. Can err: refreshed at re-sign-in. | observed |
| First user install | Earliest first-install time of a non-system app after W. Marks: U upper bound. Can err: later if that app was uninstalled. | [6] |
| Setup and restore installs | Earliest install of any package with reason setup or restore and a stored first-install time. Marks: U upper bound, and restore evidence. Can err: reason absent (0) on many installs; no first-install time on Android 10 and 12. | [5], [6] |
| Setup flags | `user_setup_complete`, `device_provisioned`. Marks: state only. Can err: no time. | [7] |
| Backup record | Ancestral token of the restore set in the `ancestral` file. Marks: restore evidence. Can err: no time; zero when no set was used. | [9] |
| Setup timestamp file | Google `initialSetupTimestamp.txt` in the backup directory. Marks: setup (where present). Can err: undocumented; observed. | observed |
| Message creator | Package stored as creator of each message. Marks: per-record provenance. Can err: names the writer, not the origin of the content; documented for sent messages, read here as the writer of any row. | [8] |
| Usage history | Oldest daily usage-statistics file. Marks: start of usage history. Can err: retention of about 10 days. | [2] |
| Recovery history | Samsung recovery history with wipe entries. Marks: wipe events. Can err: absent on some units; its own clock. | [1] |
| User record creation time | `created` in the system user's record. Marks: user creation. Can err: zero on every device here. | [10] |

### 3.2 Indicators

Table 2 lists the thirteen candidate traces. Each marks an event, and each can err in a known direction. Four mark user setup and are *upper bounds* on U: the first account add that the account manager still retains, the oldest last-password-entry time of an account, the earliest first-install time of a non-system app after W, and the earliest install with the reason "setup" or "restore". Each is the time of an event that follows setup and can be later than the true time (the account log held at most 64 rows in these extractions, a password time is refreshed at re-sign-in, an app can be uninstalled). We therefore set U to the earliest of the four, which is the tightest of the bounds and not the setup time itself.

### 3.3 Record measures

For each device and store we count records dated before W, between W and U, and from U on. The stores are the message table (row identifier and date), the call table, the media index (capture time of images and videos, where present) and the package records (install reason of apps that are not part of the system image). The unit of analysis is the device: shares are computed per device and summarised by their median and range; pooled shares are given as a second view because a few large stores dominate them. Counts of devices are exact; with 14 devices we give no intervals.

### 3.4 Labelling rules

A stored provenance label, where there is one, comes first: install reason for apps and the creator column for messages. For records without a label we test four rules that see only row identifier and date. **R1** labels a record restored if it is dated before a boundary B. **R2** labels a record restored if it is dated more than 12 hours before the highest date among records with lower identifiers; it needs no boundary. **R3** is the union of R1 and R2. **R4** closes R3 over the identifier interval it spans, because a batch is written in one burst and so occupies a contiguous range. B is W or U. All other records are labelled undetermined, not native: a record that no rule flags is consistent with having been created on the device, which is not the same as having been.

### 3.5 Synthetic fixtures

A fixture is a real SQLite table with an `AUTOINCREMENT` identifier and an integer date, filled in the order a scenario writes records: native records at their event time (a Poisson process, three per day for 180 days after W) and a restored batch of 1,500 records written in one burst at the restore time R, with creation dates spread over 365 days before R (or, for a gradual sync, 50 chunks written over 14 days between native writes). Scenarios vary the delay from W to R (0.02, 3 and 30 days), the order within the batch (ascending, descending, grouped by thread, random, or chunked), the age of the newest record in the batch (0 or 1 day before the restore), and a clock stepped back by two days for ten days. The rules see only identifier and date; the labels are kept outside the table. Each case is run 200 times with fixed seeds, and each rule is also run with the boundary wrong by −3, +3 and +14 days. The fixtures follow the documented behaviour of Section 2.1 and the batch structures observed in Section 5.6; they test the logic of the rules, not the behaviour of any restore tool.

## 4 EVIDENCE AND SETUP

### 4.1 Evidence

The Android extractions are all the full file system extractions of Android phones with a readable archive that the Lab's census of its research corpus on 27 October 2026 found; the bug reports are the Android bug reports found by the same census (Table 3).

**Table 3:** Evidence classes. All reads were read-only.

| Class | Count | Composition |
|---|---|---|
| Android full file system extractions | 14 | 12 Samsung, 2 other vendors; Android 10 to 16 (9 on Android 16, 4 on 12 to 14, fewer than 3 on 10) |
| Android bug reports | 7 | 4 Samsung, 3 from three other vendors; Android 10 to 16 |
| iOS full file system extractions (probe) | 33 | iOS 15 to 26; 30 with a message, call or photo store readable |
| Synthetic fixtures | 200 per case | SQLite tables built by the scenarios of Table 9 |

### 4.2 Setup

One runner script (Python standard library only) reads each extraction archive through its central directory, copies the few small files and databases it needs into a scratch area, reads them, and deletes the copies. It writes integers and enumerations only, and a Markdown and JSON report of aggregates from which every number here was taken. The acquisition time of an extraction is not needed, because all measures are relative to W; the file times stored in the archives, which are the source of the first boot record, were present for all 14 extractions.

## 5 RESULTS

### 5.1 Which traces exist (RQ1)

All 14 extractions carried a first boot record, a first media index, account tables with times, and both setup flags set (Table 4). Package first-install times of non-system apps after W exist on 12 devices; install reasons are recorded for at least one such app on 13, and the reason "restore" occurs on 9. The Android 10 and 12 devices store no first-install times, so the fourth upper bound of U exists on 10 devices. The backup record was present on 11 devices and held a non-zero restore-set token on 3 of them; a setup timestamp file exists on 3 devices. Messages carry a migration-tool creator on 3 devices. The oldest daily usage file reached the first boot on 0 devices (the retention of that store is about ten days [2]). The system user's creation time is zero on 14 of 14 devices. A recovery history, which can hold wipe entries [1], was read on 8 devices, 2 with a wipe entry.

### 5.2 Two boundaries, months apart (RQ2)

The first media index lies within a day of W on 12 of 14 devices and up to 540 days later on the others. The two times come from different files, so their agreement supports reading W as the first boot of the current data partition. On 2 devices, 48 non-system packages carry first-install times more than an hour before W, which is consistent with packages installed on the image before that boot; we did not test this further. The two devices where the media index lies months after W are the Android 10 and 12 devices; on one, W is a round calendar date that looks like a clock default. We regard W as unsupported on both. Without them the median gap from W to U is 69.8 days (0 to 221.8; 12 devices), at least 30 days on 9 and at least 100 days on 5.

User setup does not coincide with the first boot. Measured from W, U (the earliest of the four upper bounds) lies a median of 98 days later, between 0 and 563 days; it lies at least 30 days later on 11 of 14 devices, at least 100 days later on 7, and within a day on 2 (Fig. 2). The traces that mark setup agree with one another far better than they agree with W (Table 5): the password time and the earliest setup or restore install lie within a day of each other on 10 of 10 devices that have both, the password time and the first user install on 8 of 12, and the account add log, which keeps only the last 64 actions, agrees with the password time on 7 of 14. The Google setup timestamp, where it exists, lies 2.1 to 150.8 days after U.

**Table 4:** Prevalence of the traces of Table 2 on the 14 Android extractions.

| Trace | Devices | Of |
|---|---|---|
| First boot record (W) | 14 | 14 |
| First media index | 14 | 14 |
| Account add log | 14 | 14 |
| Account password time | 14 | 14 |
| First user install after W | 12 | 14 |
| Install reasons recorded (any non-system app) | 13 | 14 |
| of which: reason restore on at least one non-system app | 9 | 14 |
| of which: reason setup on at least one non-system app | 8 | 14 |
| Both setup flags set | 14 | 14 |
| Backup record readable / restore-set token non-zero | 11 / 3 | 14 |
| Google setup timestamp file | 3 | 14 |
| Messages written by a migration tool | 3 | 13 with messages |
| Oldest usage file reaches W | 0 | 14 |
| Recovery history read / with a wipe entry | 8 / 2 | 14 |
| System user creation time non-zero | 0 | 14 |

**Table 5:** Offset of each trace from the first boot record W, in days, over the devices that have it: median, range, number of devices within one day of W and more than 30 days after W.

| Trace | Devices | Median | Range | ≤ 1 day | > 30 days |
|---|---|---|---|---|---|
| First media index | 14 | 0.0 | 0.0 to 540.5 | 12 | 2 |
| Account add log | 14 | 221.2 | 40.3 to 2,469.6 | 0 | 14 |
| Account password time | 14 | 115.6 | 26.3 to 563.0 | 0 | 13 |
| First user install | 12 | 70.0 | 0.0 to 569.9 | 2 | 9 |
| Earliest restore install | 7 | 58.3 | 26.4 to 116.6 | 0 | 6 |
| Earliest setup install | 7 | 116.6 | 26.3 to 460.4 | 0 | 6 |
| Google setup timestamp | 3 | 190.7 | 124.0 to 207.6 | 0 | 3 |
| U (earliest of four bounds) | 14 | 98.1 | 0.0 to 563.0 | 2 | 11 |

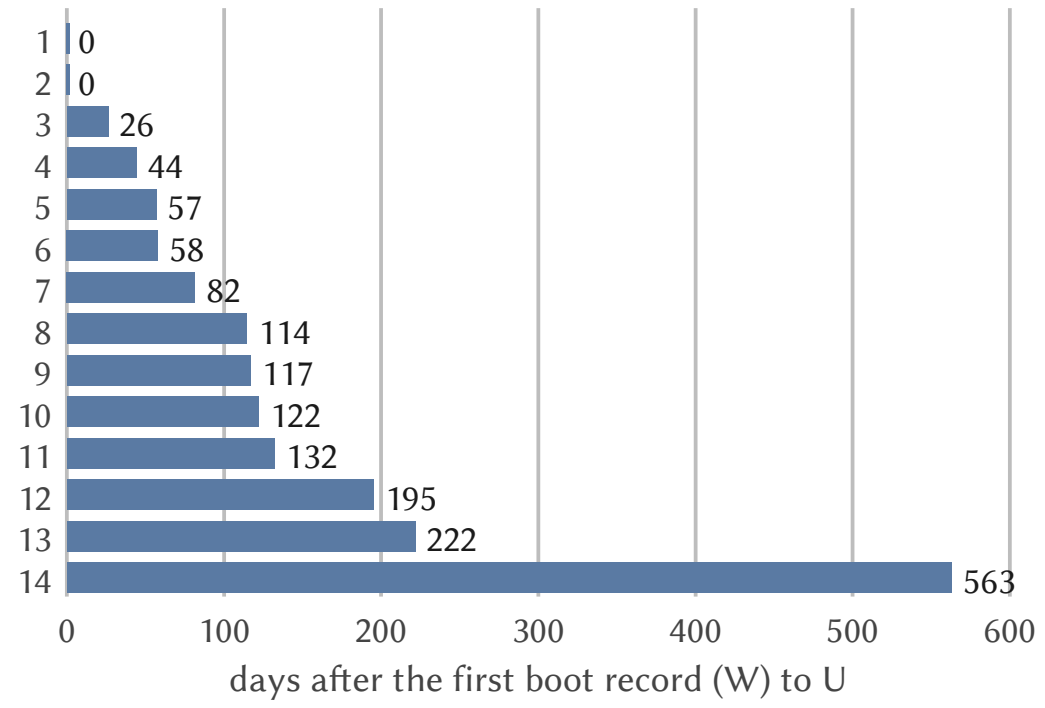

**Figure 2:** Days from the first boot record W to U for each of the 14 devices, sorted. The two shortest bars are shorter than one day.

## 5.3 What predates the boundary (RQ3)

Table 6 gives the share of records dated before W. In restored stores, messages and calls keep their original dates, so the store shows records from before the device existed: 7 of the 13 devices with messages have some, 5 a majority; calls 5 of 13 and 3. The median device holds 0.9% of its messages and 0.0% of its calls before W, but the distribution is bimodal: the share is zero on 6 of 13 message stores and above 50% on 5. Between W and U lie a further 5,820 messages and 1,726 calls, which we count as doubtful. Capture times of photographs and videos before W occur on 10 of 14 devices, with a median share of 23.8% and a range of 0.0% to 83.8%; a further 5,569 items were captured after W but entered the index more than seven days after their capture.

Install times behave differently. The package manager stamps an app with the time of installation and records the reason: of 1,302 apps outside the system image on the 14 devices, 318 (24%) carry the reason "restore", 46 "setup", 564 "user" and 374 no reason. The reason "restore" occurs on 9 devices; per device its share is 16% at the median and 51% at most. So the install date of a restored app is the date of the restore, and the original install date is not in the package records. The local state of the Play store does not keep it either: 209 of 209 restored apps that it lists have a first-download time after W, and so do all 794 of 794 listed apps.

## 5.4 Restore signals agree only partly (RQ2, RQ3)

A device that received restored data may show it in several ways: the install reason, the backup record, the creator of messages, and records dated before W. Table 7 counts them. Of the 14 devices, 11 show at least one signal and 3 none. The signals do not coincide: 9 devices have restored installs and also records before W, 0 have restored installs but no record before W, and 2 have records before W (photograph or video capture times only) but no restored installs; the backup record holds a non-zero restore-set token on 3 devices, all 3 with restored installs, while 6 of the 9 devices with restored installs have a zero token. A zero token therefore does not show that a device was set up fresh.

## 5.5 Boundary rules against the creator field (RQ4)

On the 3 devices whose messages carry a migration-tool creator, the creator is a stored label for 46,968 messages. We apply the rules of Section 3.4, which do not read the creator, and count how many labelled messages each flags and how many messages of other creators it flags. This is a comparison with a stored label, not with the origin of the content: the other creators (the default messaging app, the telecom service) may also have written restored messages. Table 8 gives the counts. With B = W the boundary rule recalls 44,650 of 46,968 tool-written messages (94% to 100% by device); the missing messages are dated between W and U. With B = U it recalls all but 1 and flags 1 message of another creator. The order rule alone, which needs no boundary, recalls 46,741 and flags 0 others; R3 adds nothing to the boundary rule here, and R4 recovers the one message left by closing over the identifier interval, with no further message of another creator flagged.

## 5.6 How restored batches are written

The records that must be restored, those dated before W, show how a batch lands in a store. We found them in 12 device-stores (messages or calls). In 9 the identifiers of these records, taken in order, have non-decreasing dates (at least 90% of adjacent pairs), in 2 the dates fall (at most 10%) and in 1 they are mixed. In 6 device-stores the records form one contiguous run of identifiers, in 6 several runs; in 4 they all precede every later record. Restores therefore write ascending, descending and chunked batches, and a rule that assumes one order will miss the others. Where no record is dated before W, the order rule flagged 584 of 23,584 records in 14 stores; one call store accounts for 583 of them and holds 592 calls between W and U, consistent with a restored batch whose dates continue past W.

## 5.7 Rules on synthetic fixtures (RQ4)

Table 9 and Fig. 4 give the recall of the four rules when the boundary is exact, for 200 runs of each scenario. The boundary rule is complete when the restore happens at setup and loses the part of the batch dated after W when the restore comes later (92.1% recall after a 30-day delay, because 29 of the 364 days the batch spans lie after W). The order rule alone depends on the order of the batch: it catches 5.0% of an ascending batch written at setup and nearly all of a descending, threaded or random one, and it is complete for a restore written after native records exist, because the first restored record then falls below the highest date already present. The union recalls at least 99.98% in every scenario. Costs appear on the native side. With a clock stepped back for ten days the order rule flags 0.7% of native records; closing over the identifier interval (R4) adds native records that lie between chunks of a gradual sync (7.6% of native records) and, after a clock step, 48.5%.

**Table 6:** Share of records dated before the first boot record W (percent), per device: devices with the store, devices with at least one such record, devices where they are the majority, median, range and pooled share. For apps the share is that of installs with the reason restore, and no date comparison is involved. Messages are the rows of the text-message table; multimedia messages were not read.

| Store | Devices | Some | Majority | Median % | Range % | Pooled % |
|---|---|---|---|---|---|---|
| Messages | 13 | 7 | 5 | 0.9 | 0.0 to 99.5 | 71.4 |
| Calls | 13 | 5 | 3 | 0.0 | 0.0 to 99.6 | 23.7 |
| Media capture | 14 | 10 | 3 | 23.8 | 0.0 to 83.8 | 72.6 |
| Restored apps | 14 | 9 | – | 16.0 | 0.0 to 51.3 | 24.4 |

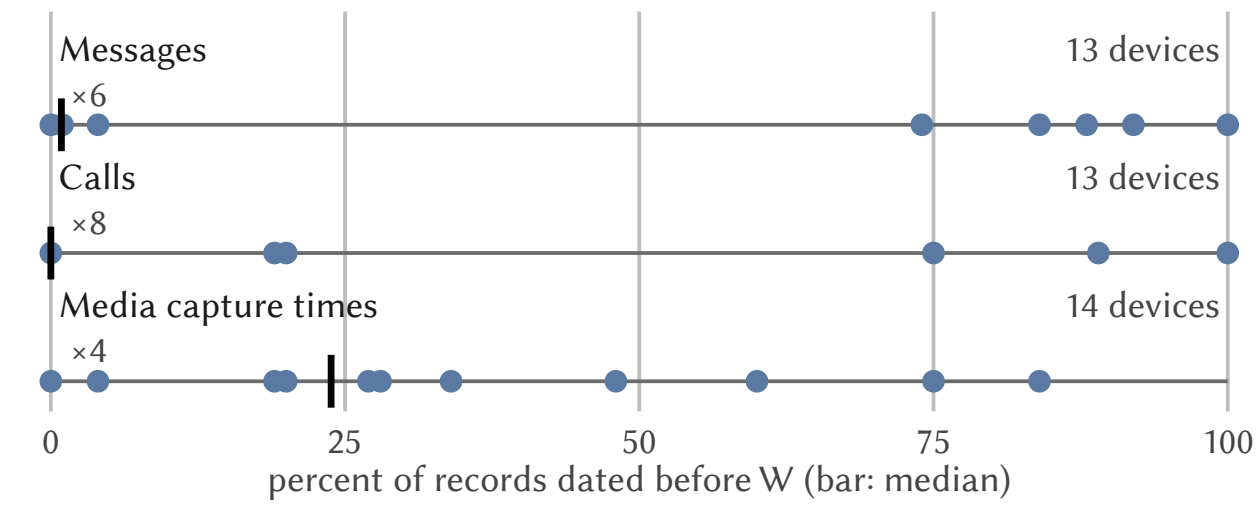

**Figure 3:** Share of records dated before W, one dot per device that holds the store.

**Table 7:** Restore signals by device (of 14 Android extractions). "Records before W" means messages, calls or photo and video capture times.

| Signal | Devices |
|---|---|
| Apps installed with reason restore | 9 |
| Backup record with non-zero restore-set token (of 11 readable) | 3 |
| Messages written by a migration tool | 3 |
| Records before W | 11 |
| At least one of the four | 11 |
| None of the four | 3 |

The boundary matters as much as the rule. With no restore (S8), a boundary placed 3 days too late flags 1.7% of native records, one placed 14 days too late flags 7.7% and a boundary 3 days too early flags none but loses 0.8% of an ascending batch written at setup (R1, S1). The error in native records is the error in the boundary divided by the length of the device's life, 180 days here. The measured distance between W and U (a median of 98 days, Section 5.2) is of the size of that whole life, which is why the choice between them decides the result.

### 5.8 Bug reports and an iOS probe

**Bug reports.** A bug report carries package install times and, in some, install reasons, but we read no first boot record, account log or message store from it. Install reasons are present in 5 of 7 reports (those of Android 15 and 16; the two Android 10 reports have none); one report lists 44 apps installed with the reason restore, all within 0.0 days of each other; the other four report none. The ancestral token printed by the backup service is zero in all 7 reports, including the one with restored installs. A bug report can therefore show that a device received a restore and when, but it does not give W, and its token does not show whether a restore took place.

**iOS probe.** We looked for files that an app creates after a restore from a backup, whose names say so (a lock file in the backup cache, and one-time tasks named for the restore launch), and took the earliest of their file times as the marker time M. Apple does not document these files, so M is an observed indicator and not a measurement of the restore. 29 of the 33 iOS extractions carry a marker. On the marker devices, messages dated more than an hour before M exist on 19 of 29 devices with messages, calls on 12 of 28 and photograph creation dates on 26 of 28. They are the majority of messages on 12 devices, of calls on 1 and of photographs on 17. The share of messages before M has a median of 21.2% and ranges from 0.0% to 99.3%; for calls the median is 0.0% and for photographs 53.0%. Like the Android stores, iOS stores can show records older than the marker, and the share varies from none to nearly all.

## 6 DISCUSSION

### 6.1 Two boundaries answer two questions

The first boot record answers "could this record have been created on this device?". A record dated before it, if the clock was right at that boot, could not. It is the stricter boundary: on the 13 devices that hold them, 165,778 messages and calls are dated before it (of 244,053), and unless the clock was wrong they were carried to the device. The user setup boundary answers "was this record probably restored?". It is later by a median of 98 days, and the restored part of a batch lies in that gap: against the creator field, the first boot boundary missed 0% to 6% of the tool-written messages and the setup boundary all but 1. Which boundary to use depends on the claim. To exclude a record from the device's own history with certainty, use W. To label records as restored without missing the tail of a batch, use U, and accept that records created at a store or in a test before setup may be labelled with them.

### 6.2 Wording a claim about a date

An examiner asked whether something happened on date D should check D against W and U first. If D is before W, no record on this device can show it happened on D, however old the records look; if D lies between W and U, records dated D are doubtful; if D is after U, records dated D may be taken as the device's own unless a provenance field or a rule says otherwise. For app installs the question needs a different answer: the install date of an app with the reason restore is the date of the restore (Section 5.3), so "was the app installed on D" is answered by the restore date, and "did the user use the app on D" needs evidence from the app's own data.

### 6.3 What the measurements do not show

A gap between W and U does not say why it exists; a device may have booted at a factory, a depot or a shop, or after a reset by its user, and the traces cannot tell these apart. The restore signals are evidence of a restore, not a measure of how much was restored. A record that no rule flags is consistent with having been created on the device, not shown to have been.

## 7 THREATS TO VALIDITY AND LIMITATIONS

**Construct.** W is a file time written by the platform at the first boot-time record; it is wrong if the clock was wrong at that boot, and any later wipe replaces it. We cannot check the clock at that boot independently; the agreement of the first media index with W on 12 of 14 devices supports the reading, and 48 non-system packages on 2 devices pre-date it by more than an hour. U is an upper bound, not the setup time. Records are called predating when dated more than an hour before a boundary; a record with a wrong date, for example from a message with a bad network time, would be labelled wrongly. The creator field names the writer of a message, not where the content came from, and we classed one package name as a migration tool from its name and its public app listing without verifying it against vendor documentation.

**Internal.** The comparison with the creator field covers 3 devices and one migration tool; U was computed from account and package records and not from message dates, but the same devices were used to design the question. We read the devices as acquired and used no record of the restore method or its date, so the restore time R is inferred from install reasons, and the structure of restored batches (Section 5.6) is observed, not controlled. The fixtures encode our reading of the documented behaviour; they show the logic of the rules, not how any restore tool writes.

**Table 8:** Rules against the creator field, 3 devices pooled: 46,968 messages written by the migration tool and 2,707 messages of other creators. Recall is the share of tool-written messages flagged; the last column counts flagged messages of other creators.

| Rule | Tool-written flagged | Recall % | Other flagged |
|---|---|---|---|
| R1, B = W | 44,650 | 95.06 | 0 |
| R1, B = U | 46,967 | 100.00 | 1 |
| R2, order only | 46,741 | 99.52 | 0 |
| R3, B = U | 46,967 | 100.00 | 1 |
| R4, B = U | 46,968 | 100.00 | 1 |

**Table 9:** Fixture results with an exact boundary: recall of restored records / share of native records flagged, percent, mean of 200 runs. "At setup", "day 3" and "day 30" give the restore time after W; clock step: back by two days for ten days. S8 and S9 hold no restored records.

| Scenario | R1 | R2 | R3 | R4 |
|---|---|---|---|---|
| S1 at setup, ascending | 100.0 / 0.0 | 5.0 / 0.0 | 100.0 / 0.0 | 100.0 / 0.0 |
| S2 at setup, descending | 100.0 / 0.0 | 99.8 / 0.0 | 100.0 / 0.0 | 100.0 / 0.0 |
| S3 at setup, by thread | 100.0 / 0.0 | 98.0 / 0.0 | 100.0 / 0.0 | 100.0 / 0.0 |
| S4 at setup, random | 100.0 / 0.0 | 99.4 / 0.0 | 100.0 / 0.0 | 100.0 / 0.0 |
| S5 day 3, ascending | 99.4 / 0.0 | 100.0 / 0.0 | 100.0 / 0.0 | 100.0 / 0.0 |
| S6 day 30, ascending | 92.1 / 0.0 | 100.0 / 0.0 | 100.0 / 0.0 | 100.0 / 0.0 |
| S7 day 30, by thread | 92.1 / 0.0 | 100.0 / 0.0 | 100.0 / 0.0 | 100.0 / 0.0 |
| S11 gradual sync, 14 days | 100.0 / 0.0 | 98.9 / 0.0 | 100.0 / 0.0 | 100.0 / 7.6 |
| S8 no restore | – / 0.0 | – / 0.0 | – / 0.0 | – / 0.0 |
| S9 no restore, clock step | – / 0.0 | – / 0.7 | – / 0.7 | – / 0.7 |
| S10 setup, ascending, clock step | 100.0 / 0.0 | 7.0 / 0.6 | 100.0 / 0.6 | 100.0 / 48.5 |

**External.** The sample is one collection of 14 Android extractions, 12 Samsung and 9 on Android 16, from the Lab's research corpus; it is not a population of seized phones, and each vendor other than Samsung is represented by fewer than three devices. We read no first boot record from the seven bug reports. The iOS probe rests on file names that Apple does not document and on marker times that an app may refresh. Messages, calls, media and package records were studied; contacts, MMS and the photo library on Android were not.

**Limitations.** The study contains no restore experiment, so no claim here describes what a restore tool does; only what stored data show. The corpus runner is not released (Artifact Availability).

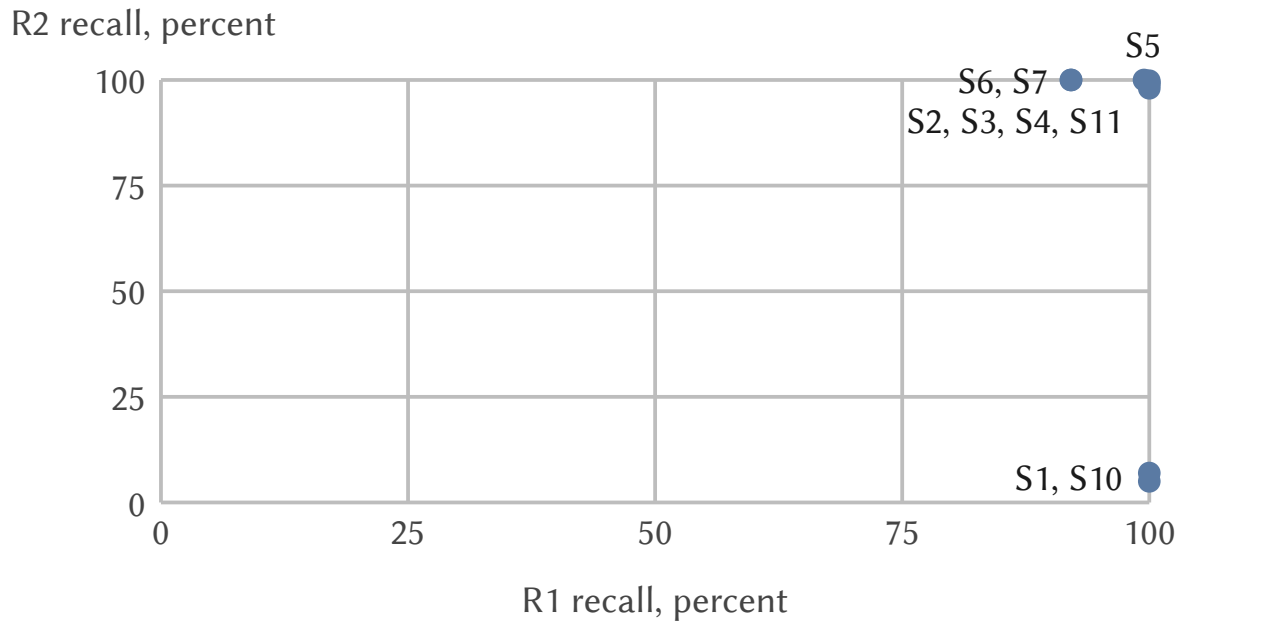

**Figure 4:** Recall of the order rule R2 against that of the boundary rule R1 for the nine scenarios with a restore (Table 9; exact boundary). No scenario is low on both axes, which is why their union R3 is complete.

**Table 10:** A ladder for the claim a dated record supports. Each rung needs the evidence in the middle column; a lower rung is used when a higher one fails.

| Claim | Evidence needed | What the date then shows |
|---|---|---|
| Created on this device | Provenance field names a native writer, or the record is dated after U and no rule flags it | The date may be taken as that of an event on this device |
| Probably restored | Dated between W and U, or flagged by the order rule | The date is that of another handset or of the backup; this device did not exist or was not in use |
| Restored | Dated before W; or install reason restore; or a migration tool as creator | The date is from elsewhere; for installs it is the date of the restore |
| Undetermined | No boundary, no provenance field | Say so; do not read the date as the device's own |

## 8 CONCLUSION AND FUTURE WORK

A phone has two starts that an examination can measure: the first boot of its data partition, which every extraction here recorded, and the first evidence that a user set it up, which came a median of 98 days later. The password time agrees within a day with the earliest setup or restore install on all 10 devices that have both and with the first user install on 8 of 12, while the retention-limited account log agrees with it on 7 of 14; all of them disagree with the first boot by months. Restored data show on both sides of the first boot: 7 of the 13 devices with messages hold messages dated before it, restored installs make up 24% of the installs outside the system image, and the install date of a restored app is the date of the restore. A boundary placed at user setup separated the 46,968 messages that a migration tool wrote from the 2,707 of other creators with 1 missed and 1 flagged, where the first-boot boundary missed up to 6% of them; order rules need no boundary but depend on how a batch was written.

A report should name the boundary it used and say "undetermined" when a record has neither a boundary nor a provenance field to support it.

Open work: controlled restores on a Lab-owned device with known methods, to link the observed batch structures to the tools that write them; the same measures on more vendors and on the iOS stores whose markers are documented by controlled test; contacts, MMS and the photo library; and the setup timestamp file, which we saw on three devices and could not document.

## ARTIFACT AVAILABILITY

The fixture generator, the four labelling rules and the evaluation behind Table 9, Figure 4 and the boundary-error statements of Section 5.7 are released at https://github.com/danjethh/sectratek-lab-artifacts/tree/main/reports/lab-tr-0028 and give those values from a fixed seed. Also released are read-only readers that turn an extraction archive, a bug report or an iOS archive into lists of integers (paths come from a configuration file), the analysis that computes Tables 4 to 8 from them, a generator of invented archives and a self-test (Python, standard library only; code MIT, data and text CC BY 4.0). On invented archives these show that the pipeline runs; they do not reproduce the findings. Tables 4 to 8, Figures 2 and 3 and Sections 5.1 to 5.6 and 5.8 rest on the Lab's research corpus, which cannot be released. The readers of two vendor-specific stores are not released.

## DATA HANDLING STATEMENT

The evidence for this study is drawn from the Lab's research corpus of mobile device extractions and bug reports and was read only, together with synthetic data. The analysis read stored extractions and bug reports only. Message and call stores were read for dates, row identifiers and numeric codes, and for the package name stored as the creator of a message; the media index for times and type codes; the package records for package names, times and install reasons; the Play app-state and library tables for package names, times and type codes; account stores for row identifiers, action codes and times; the backup record for its header. The report gives aggregate counts and device classes. It contains no identifiers, personal data, message or file content, or case information. The device counts rest on hash comparisons of device identifiers recorded in the Lab's private notes: the 14 Android extractions are 14 phones, the 33 iOS extractions analysed are 33 phones, and the seven bug reports come from seven devices (six distinct serial-number hashes; the seventh report differs in manufacturer). Whether a bug report and an extraction come from the same phone was not tested; the two sets are never combined in a count.